\documentclass[prl,twocolumn,showpacs,superscriptaddress]{revtex4}

\usepackage{graphicx}
\usepackage{amsmath}
\usepackage{amssymb}
\usepackage{epsfig}

\usepackage{ctable}
\usepackage{slashbox}

\renewcommand{\v}[1]{{\bf #1}}

\def\eqa{\begin{eqnarray}}
\def\eea{\end{eqnarray}}
\def\be{\begin{eqnarray}}
\def\ee{\end{eqnarray}}
\newcommand{\nn}{\nonumber\\}

\newcommand{\Del}{\Delta}
\newcommand{\eps}{\epsilon}

\newcommand{\Ga}{\Gamma}

\newcommand{\La}{\Lambda}

\usepackage{color}

\begin{document}

\title{Monolayer NbF$_4$: a $4d^1$-analogue of cuprates}

\author{Yang Yang}
\affiliation{Texas Center for Superconductivity and Department of Physics, University of Houston, Houston, Texas 77204, USA}
\affiliation{College of Physics and Electronic Engineering, Zhengzhou University of Light Industry, Zhengzhou, 450002, China}

\author{Wan-Sheng Wang}
\affiliation{Department of Physics, Ningbo University, Ningbo 315211, China}

\author{C. S. Ting}
\affiliation{Texas Center for Superconductivity and Department of Physics, University of Houston, Houston, Texas 77204, USA}

\author{Qiang-Hua Wang}
\email{qhwang@nju.edu.cn}
\affiliation{National Laboratory of Solid State Microstructures $\&$ School of Physics, Nanjing
University, Nanjing, 210093, China}
\affiliation{Collaborative Innovation Center of Advanced Microstructures, Nanjing University, Nanjing 210093, China}

\begin{abstract}
The electronic structure and possible electronic orders in monolayer NbF$_4$ are investigated by density functional theory and functional renormalization group. Because of the niobium-centered octahedra, the energy band near the Fermi level is found to derive from the $4d_{xy}$ orbital, well separated from the other bands. Local Coulomb interaction drives the undoped system into an antiferromagnetic insulator. Upon suitable electron/hole doping, the system is found to develop $d_{x^2-y^2}$-wave superconductivity with sizable transition temperature. Therefore, the monolayer NbF$_4$ may be an exciting $4d^1$ analogue of cuprates, providing a new two-dimensional platform for high-$T_c$ superconductivity.
\end{abstract}





\maketitle

{\em Introduction}: The search for high-$T_c$ superconductors (HTS) has lasted for decades
after the discovery of the cuprates\cite{cuprates}. A natural idea is to
search for structural and electronic analogue of the cuprates, characterized by layered structure and a single energy band near the Fermi level derived from the $d$-orbital of the transition element.
The correlation effect on top of the quasi two-dimensional (2D) band structure should make HTS very likely
\cite{Arita1999,Arita2000,Monthoux1999,Monthoux2001,Sakakibara2010,Sakakibara01,Sakakibara02}.

Recently, the infinite-layer nickelate NdNiO$_2$ has been discovered to be a superconductor
with transition temperature $T_c=9\sim15$~K upon doping\cite{nickelate}. As Cu$^{2+}$ in cuprates, Ni$^+$ is in the
$3d^9$ configuration, with $3d_{x^2-y^2}$ partially occupied. However, the hybridization with
Nd-$5d$ states near the Fermi level increases the complexity and possibly also limits the transition temperature.

Meanwhile, the development in 2D materials provides another avenue to find unconventional superconductors. Since the the discovery of graphene\cite{graphene1,graphene2}, many 2D materials have been explored for superconductivity (SC),
such as doped phosphorene, single-layer B$_2$C, and 2D boron, {\it etc}.\cite{phosphorene,B2C,boron1,boron2}
Specifically, one of the most widely studied 2D superconductors is 2H-NbSe$_2$, where SC develops within the
charge-density wave (CDW) phase\cite{Xi2015}. It becomes more encouraging to search for new 2D superconductors after the
observation of SC in the twisted bilayer graphene\cite{magic1,magic2} and high $T_c$ SC in
the one-unitcell monolayer Bi$_2$Sr$_2$CaCu$_2$O$_{8+\delta}$\cite{ZhangYB}.

In this Letter we try to find a close 2D analogue of cuprates with non-copper transition element. The material we consider is the monolayer NbF$_4$, which has niobium-centered fluorine octahedra mimicing the oxygen octahedra in cuprates.
By density functional theory (DFT) studies, we find a single band with dominant $4d_{xy}$ character near the Fermi level, while the other bands are well separated. Upon inclusion of the local Coulomb interaction, the parent compound is found to be an anteferromagnetic insulator within DFT theory. By functional renormalization group (FRG) calculation on the basis of an effective Hubbard model, we find  $d_{x^2-y^2}$-wave SC upon electron/hole doping. These features make NbF$_4$ a promising and close analogue of cuprates, and may act as a new 2D platform for the exploration of HTS.

\begin{figure}
\includegraphics[width=8.5cm]{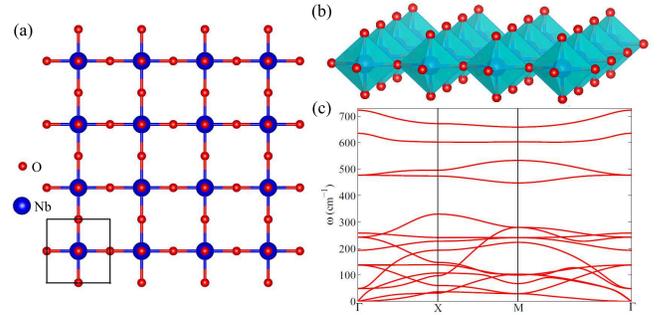}
\caption{(Color online) Crystal structures of monolayer NbF$_4$ from top view (a) and side view (b). The black solid lines denote the unit cell. (c)
Phonon dispersion of monolayer NbF$_4$.}\label{cry}
\end{figure}

{\em Crystal and DFT results}:
We start from the crystal structure of monolayer
NbF$_4$. The crystallographic parameters are taken from Ref.\cite{NbF4cry}.
The top view and side view of the crystal structure are shown in
Fig.\ref{cry}(a) and (b), respectively.
The Nb and F atoms form a square lattice
and each Nb atom is centered in fluorine octahedra, resembling the
oxygen octahedra surrounding copper ions in cuprates.
The DFT calculations are performed using
the Quantum ESPRESSO (QE) package\cite{qe}.
The projected augmented wave (PAW) pseudopotentials with
generalized gradient approximation (GGA)
are adopted for the exchange-correlation energy\cite{GGA}.

The structures are fully relaxed until the force on
each atom is $< 0.001$~eV/\AA. After relaxation, the bond lengths
between the Nb and F atoms is 2.07~\AA (inplane) and 1.87~\AA (apical).
For comparison, in cuprate the  planer Cu-O bond length 1.96~\AA~and
apical length 2.30~\AA\cite{length}. The fluorine octahedra can be seen as the oxygen octahedra rotated by 90 degrees.
To confirm the thermodynamical stability of monolayer NbF$_4$,
the phonon dispersion is further calculated. Fig.\ref{cry}(c) shows that no imaginary frequencies are observed
in the phonon dispersion, indicating good kinetic stability of monolayer NbF$_4$. Besides,
previous calculations have shown that the monolayer NbF$_4$ may be easily exfoliated from their parent compounds\cite{NbF4cry}, encouraging the study of the monolayer.

\begin{figure}
\includegraphics[width=7.5cm]{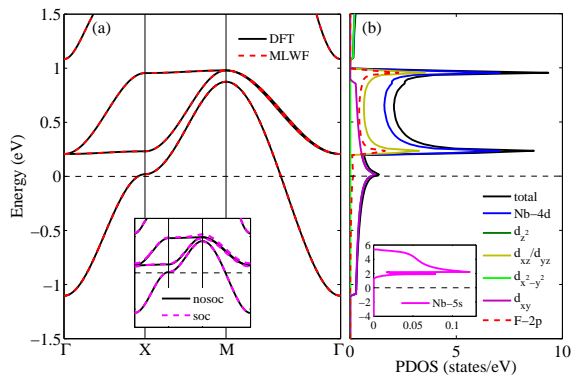}
\caption{(Color online) (a) Band structure obtained by DFT calculations
(solid black lines) and MLWF fitting (dashed red lines) for NM state. The inset shows the comparison of band structure with and without SOC effect.
(b) The total and partial DOS for monolayer NbF$_4$ in the NM state. The inset shows the PDOS of Nb-$5s$ orbital.}
\label{band}
\end{figure}

\begin{figure}
\includegraphics[width=8.5cm]{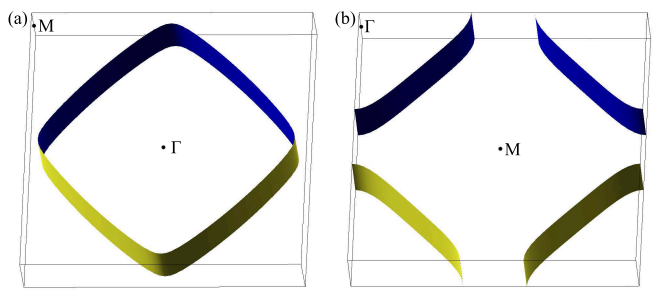}
\caption{(Color online) Fermi surface sheets of monolayer
NbF$_4$ for NM state with $E_F=0$ (a) and
$E_F= 0.1$eV (b), respectively.}
\label{ldafs}
\end{figure}

We continue to discuss the electronic structure of monolayer NbF$_4$.
As shown in Fig.\ref{band}(a),
there is only a single band crossing the Fermi level $E_F$.
The partial density of states (PDOS) is shown in Fig.\ref{band}(b). We see that the
Nb-$4d_{xy}$ plays a dominant role to the electronic states near $E_F$. In fact there is no apparent
hybridization with the other $4d$ orbitals. On the other hand, the bands derived from F-$2p$ are rather flat and mainly located -5.7 eV below the Fermi level (not shown), indicating negligible hybridization between F-$2p$ and  Nb-$4d$ orbitals. The inset of Fig.\ref{band}(b) shows
the PDOS of Nb-$5s$, which is above $E_F$ hence unoccupied. Since the atomic configuration of Nb is $4d^4 5s^1$, intuitively the four electrons in the $4d$-orbitals
would be transferred to the four surrounding F$^-$ ions.
However, the PDOS shows that one electon in $5s$- and three electrons in $4d$-orbitals are lost instead,
resulting in a half-filled $4d_{xy}$ orbital. This is in fact a natrual effect of the crystal field: The compressed octahedra splits the $4d$ levels in the ascending order of $d_{xy}$, $d_{xz, yz}$, $d_{x^2-y^2}$ and $d_{3z^2-r^2}$, and also pushes up the $5s$ level, leaving $d_{xy}$ the lowest level in the $4d$ and $5s$ levels. Therefore the Nb atom settles in the configuration close to $4d^1$.
We also performed full-relativistic DFT calculations including spin-orbit coupling (SOC) effect. The comparison between the band structures with and without SOC are shown in the inset of Fig.\ref{band}(a).
No significant difference is found, especially near the Fermi level. Thus we shall not discuss the SOC
effect henceforth for brevity.

The band near $E_F$ with Nb-$4d_{xy}$ character gives rise to
remarkably simple FS sheets, as seen in Fig.\ref{ldafs}(a).
A rounded-square FS around $\Gamma$ point cuts the the Brillouin zone into two parts of
roughly the same size. Thus the FS are featured by a quasi-nesting vector $(\pi,\pi)$.
Since this system is near half filling, the FS can evolve into either
electron pockets around $\Gamma$ point or hole pockets around M point
by rigid band shift, resulting in Lifshitz transition. As an example, Fig.\ref{ldafs}(b) shows the FS evolves into a hole pocket around the M point when $E_F$ is shifted upward by $0.1$eV.
There is a Van Hove singularity (VHS) in the vicinity of X point of the band dispersion, leaving
a peak in DOS near $E_F$. The proximity to the VHS is similar to that in cuprates, making the system highly susceptible to electron instabilities when the interaction betweeen electrons is taken into account.

\begin{figure}
\includegraphics[width=8.5cm]{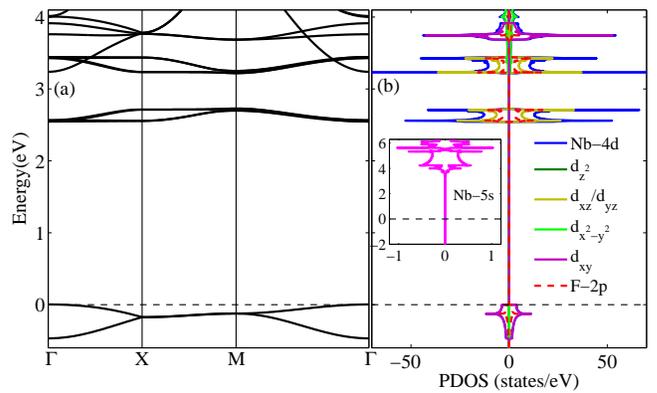}
\caption{(Color online) The band (a) and PDOS (b) of monolayer NbF$_4$ for G-type AFM state within GGA$+U$.
The inset shows the PDOS of Nb-$5s$ orbital in this case.}
\label{Udos}
\end{figure}

Indeed, we obtain the magnetic ground state by employing DFT. The calculations
are performed within both 
$2\times2\times1$ and $3\times3\times1$ supercells.
We consider the nonmagnetic (NM) state, ferromagnetic (FM) state,
G-type antiferromagnetic (AFM) state, and C-type AFM state.
We find the G-type AFM state is energetically favored,
about 131.7 meV (per Nb atom) lower than the NM
state. The magnetic moment estimated for the AFM state is about 0.38 $\mu B$ per Nb.
Including the local coulomb interaction $U$, the G-type AFM state becomes
426.6 meV lower than the NM state, and the magnetic moment increases to
about 0.44 $\mu B$.

We further perform GGA$+U$ calculation to incorporate the correlation effect better.\cite{GGAU} It is known that if the on-site Coulomb interaction is included in the AFM
configurations, the electronic state of cuprates can become insulating. Instead, without the local interaction, the ground state is metallic, as demonstrated in recent studies of Ba$_2$CuO$_{3+\delta}$\cite{Ba2CuO3}.
In our calculation, we determine $U\sim 4.05$~eV using density-functional perturbation theory (DFPT) as implemented in
QE\cite{calU}, which is close to the former results\cite{cRPA1,cRPA2}.
With this value of $U$ in the G-type AFM
configuration, the obtained bands and PDOS are shown
in Fig.\ref{Udos}(a) and (b), respectively. Obviously,
the previous nonmagnetic metallic
state now becomes AFM insulating. Differently from cuprates,
the top of the valence band is located at the $\Gamma$ point, and the band gap
is $\sim 2.5$ eV, larger than that
of Ca$_2$CuO$_3$ (1.7 eV) and Sr$_2$CuO$_3$ (1.5 eV)\cite{gap}.
Owing to the $4d^{1}$ configuration,
the PDOS of $4d_{xy}$ is splitted into two parts while the other four $4d$ orbitals are
almost unaffected. Additionally, the Nb-$5s$ remains unoccupied but more
electrons are transferred from Nb-$4d_{xy}$ to F-$2p$, resulting in slightly stronger $p-d$ hybridization, as seen from the PDOS peaks of $4d$ and F-$2p$ in Fig.\ref{Udos}(b).

Since only Nb-$4d_{xy}$ orbital dominates the low energy physics in monolayer NbF$_4$, it
is rather easy to derive a single-orbital tight-binding model.
Firstly, we obtain an effective model including five Nb-$4d$ orbitals
by the maximally-localized Wannier function method\cite{wannier1,wannier2}.
Then we obtain an effective single-orbital model to
describe the electronic properties near $E_F$. The band dispersion in momentum space can be well approximated by
\eqa  \eps_\v k &=& 2t_1(\cos k_x+\cos k_y)+4t_2\cos k_x \cos k_y \nn
           &+& 2t_3[\cos(2k_x)+\cos(2k_y)] \nn
           &+& 4t_4[\cos(2k_x)\cos k_y+\cos k_x \cos(2k_y)]-\mu, \label{HTB} \eea
where $t_1=-0.245$~eV, $t_2=-0.017$~eV, $t_3=0.002$~eV, $t_4=-0.001$~eV, and the chemical potential $\mu=0.054$~eV in the undoped case. Note the hopping integral here follows from the direct overlap between $d_{xy}$ orbitals, without having to be bridged by the $p_{x, y}$-orbitals (which would be necessary for hopping between $d_{x^2-y^2}$ orbitals). This results in a relatively narrow band, of width $\sim 2$~eV, crossing $E_F$.

{\em FRG analysis}: The similarity in the electronic structures of monolayer NbF$_4$ and cuprate motivates us to consider the possibility of superconductivity in monolayer NbF$_4$. We consider the Hubbard model with the normal state described by Eq.\ref{HTB} and a local Hubbard interaction $U$.
The interaction can lead to competing collective fluctuations that must be treated on equal footing. For this purpose, we use the singular-mode
functional renormalization group (SMFRG) method\cite{wws1,xyy1,xyy2,wws2,xyy3,yy,sro,yy2,wws3,wws4}.
In this method, the one-particle-irreducible four-point interaction vertex function $\Ga$ is calculated iteratively versus a decreasing
energy scale $\La$. During the FRG flow,
we decompose $\Ga$ into scattering matrices in the basis of fermion bilinears
(separately) in the pairing (SC),
spin-density-wave (SDW) and CDW channels.
The divergence of the negative leading eigenvalue (NLE) $S$ of the scattering matrices signals an emerging order at the associated collective momentum, and the internal structure of the order parameter is described by the associated eigenfunction of the scattering matrix. The technical
details can be found elsewhere\cite{wws1,xyy1,xyy2,wws2,xyy3,yy,sro,yy2,wws3,wws4}.

\begin{figure}
	\includegraphics[width=8.5cm]{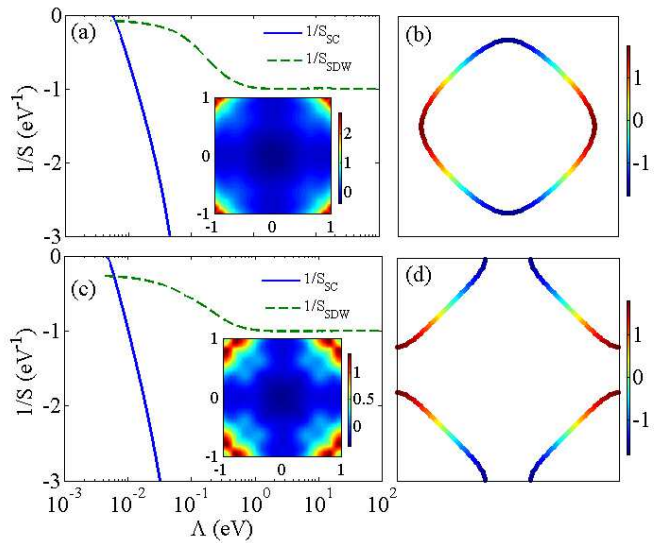}
	\caption{(Color online)(a) FRG flow of 1/$S_{\text{SC,SDW}}$ versus the running
		energy scale $\La$ for $n=0.85$ and $U=1.0$~eV. The inset shows $|S_{\text{SDW}}(\v q)|$ in the Brillouin zone at the final energy scale. (b) The Fermi surface
		and gap function $\Del(\v k)$ (color scale). (c) (d) shows the FRG flow and gap funtion $\Del(\v k)$ for n = 1.26 and $U=1.0$~eV.}
	\label{frg}
\end{figure}

We have performed systematic calculations by varying the filling level $n$ and
the local Coulomb interaction $U$.
The maximum value of $n$ is restricted ($n<1.4$) to avoid
the influence of the other $4d$ orbitals.
We select two typical cases for illustration.

For hole doping with $n = 0.85$ and $U=1$~eV, the FRG flow versus $\La$ is
shown in Fig.\ref{frg}(a). Since the CDW channel remains weak during the flow,
we shall not discuss it henceforth.
From the flow, we find the SDW channel is enhanced in the intermediate stage, but becomes flat at
low-energy scales because of lack of phase space for low-energy particle-hole excitations.
The NLE of SDW channel $S_{\text{SDW}}$ is associated with scattering momentum $\v Q =(\pi,\pi)$ and remains
almost unchanged during the flow. The inset of Fig.\ref{frg}(a) shows
$S_{\text{SDW}}(\v q)$ versus $\v q$ at the final stage, which is peaked around $(\pi,\pi)$. We checked that this scattering mode describes site-local spins, indicating strong AFM fluctuations consistent with
the DFT results.
Triggered by such spin fluctuations, the pairing interaction $S_{\text{SC}}$ are
enhanced and diverges at the final energy scale, indicating the system develops a SC instability.
The pairing function, or the eigenfunction of the Cooper scattering eigen mode, is found to be
$\cos k_x-\cos k_y$ with $d_{x^2-y^2}$-wave symmetry. This symmetry is made explicit by the plot of the pairing function on the Fermi surface in Fig.\ref{frg}(b).

Similar calculations are performed for electron doping with $n = 1.26$ and $U =1.0$~eV. The FRG flow and pairing gap on FS
are presented in Fig.\ref{frg}(c) and (d), respectively.
The leading scattering momentum in the SDW channel is $(\pi,\pi)$ at high-energy scales and changes
slightly at lower energy scale owing to the change of the FS topology.
At the final stage, the $S_{\text{SDW}}(\v q)$ peaks
at $\v Q=(0.94,0.81)\pi$, which is also near $(\pi,\pi)$, as shown in the inset of Fig.\ref{frg}(d).
In the pairing channel, the pairing symmetry remains to be $d_{x^2-y^2}$. This is because the attractive pairing interaction is triggered by SDW fluctuations in the intermediate energy window, while the change of FS topology is important only for low-energy particle-hole excitations.

\begin{figure}
\includegraphics[width=8.5cm]{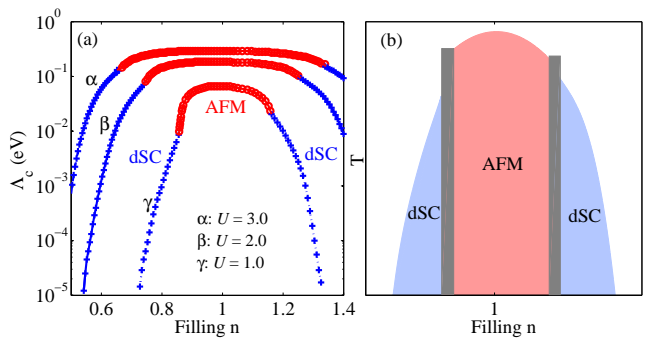}
\caption{(Color online)(a) The FRG diverging energy scale $\La_c$ plotted as
a function of filling level n.(b) A schematic temperature-doping phase diagram near
n=1. The gray region denotes the transition between SC and AFM. }
\label{pd}
\end{figure}

In Fig.\ref{pd}(a) we plot the divergence scale (representative of the transition temperature) in the AFM or SC channel, whichever is larger, as
a function of the filling level $n$, for three values of $U$. The AFM order is favored near half filling ($n=1$), and a larger $U$ enlarges the AFM region significantly. The SC order is favored upon small electron/hole doping. We observe that near the phase boundary between AFM and SC, the divergence scale is sizable, ranging from $8$~meV at $U=1$~eV, to $100$~meV at $U=3$~eV. This suggests that the monolayer NbF$_4$ may be a HTS.
On the basis of the above results, we draw a schematic phase diagram, Fig.\ref{pd}(b), for monolayer NbF$_4$ with a realistic local Coulomb interaction, of the order of the values we considered. The grayed region indicates the transition between AFM and SC.

We should remark that since the $d_{xy}$-derived band is relatively narrow, it is likely that the system is close to or in the Mott limit. In that case, the SC regime in Fig.\ref{pd}(b) may become a dome, like that in cuprates. The Mott limit is unfortunately beyond the realm of FRG on the basis of itinerant normal state. Instead of the Hubbard model, a better starting point is the $t$-$J$ model, which is however even more difficult to tackle reliably, although the same type of model has been studied extensively in the context of cuprates. We stress that our FRG within the Hubbard model provides qualitatively reliable result from weak to moderate coupling, and we leave the Mott limit in future studies.

{\em Summary}: The electronic structure and superconductivity of monolayer NbF$_4$ are investigated by DFT and FRG.
A single band derived from Nb-$4d_{xy}$ orbital is found near the Fermi level.
Near half filling the system is in the AFM state, while electron/hole doping introduces $d$-wave SC, potentially with high $T_c$. The structural and electronic similarity to that in cuprates
makes monolayer NbF$_4$ an excellent $4d^1$ analogue of cuprates and a new platform to explore HTS.

\acknowledgments{ {\em Acknowledgments}: The project was supported
by the National Key Research and Development Program of China (under Grant No. 2016YFA0300401), the National Natural Science Foundation of China (under Grant Nos.11604303, 11604168 and 11574134),
the Texas Center for Superconductivity at the University of Houston and
the Robert A. Welch Foundation (Grant No. E-1146).
YY acknowledges support by China Scholarship Council
(under Grant No. 201909440001).}

\end{document}